\documentclass[12pt]{article}

\def \l{\lambda}

\def \x{\xi}

\def \ps{\psi}

\def \G{\Gamma}

\def \La{\Lambda}

\def \la#1{\label{#1}}
\def \ift{\infty}
\def \le{\left}
\def \ri{\right}
\def \da{\dagger}

\def \lb{\lbrack}
\def \rb{\rbrack}

\def \nn{\nonumber}
\def \Tr{{\rm Tr} \,}
\newcommand \beq{\begin{eqnarray}}
\newcommand \eeq{\end{eqnarray}}
\newcommand \bea{\begin{eqnarray*}}
\newcommand \eea{\end{eqnarray*}}
\newcommand \ben{\begin{enumerate}}
\newcommand \een{\end{enumerate}}
\newcommand \ba{\begin{array}}
\newcommand \ea{\end{array}}

\begin{document}

\begin{center}
{\LARGE \textbf{String Bit Models of}} \\[0pt]
\vspace{.3cm} {\LARGE \textbf{Two-Dimensional Quantum Gravity}} \\[0pt]
\vspace{.3cm} {\LARGE \textbf{Coupled with Matter}} \\[0pt]
\vspace{.7cm} {\large \textbf{C.-W. H. Lee}$^{a,b,}$\footnote{{\large e-mail
address: lee@math.uwaterloo.ca}} and \textbf{R. B. Mann}$^{a,}$\footnote{%
{\large e-mail address: mann@avatar.uwaterloo.ca}}} \\[0pt]
\vspace{.7cm} $^a$ \textit{Department of Physics, Faculty of Science,
University of Waterloo, Waterloo, Ontario, Canada, N2L 3G1.} \\[0pt]
$^b$ \textit{Department of Pure Mathematics, Faculty of Mathematics,
University of Waterloo, Waterloo, Ontario, Canada, N2L 3G1.} \\[0pt]
\vspace{.4cm} {\large July 24, 2003} \\[0pt]
\vspace{.7cm} {\large \textbf{Abstract}}
\end{center}

\noindent We extend the formalism of Hamiltonian string bit models of
quantum gravity type in two spacetime dimensions to include couplings to
particles. We find that the single-particle closed and open universe models
respectively behave like empty open and closed universes, and that a system
of two distinguishable particles in a closed universe behaves like an empty
closed universe. We then construct a metamodel that contains all such
models, and find that its transition amplitude is exactly the same as the $%
sl(2)$ gravity model.

\vspace{.5cm}

\begin{flushleft}
\end{flushleft}
\pagebreak

\section{Introduction}

\label{s1}

Two-dimensional gravity continues to be a subject of lively interest, in
large part because of the hope that its simplified setting offers hope for
developing approaches to quantum gravity that can be extended to higher
dimensions. One such approach has been that of discretizing spacetime in
terms of random triangulations. These discrete models have provided analytic
insights that are currently out of reach of continuum methods. For example
incorporation of spacetimes of arbitrary topology can be attained via the
double scaling limit (See refs.\cite{dgz} and \cite{adj} and the citations
therein).

One promising approach recently introduced has been that of Hamiltonian
string bit models \cite{0108149}. Originally introduced as a means of
regulating string theory \cite{gt, thorn, ks}, in these models the spatial
metric degree of freedom is discretized by introducing a distance cutoff $a>0
$. Equal-time slices of spacetime are taken to be polygonal loops of volume $%
na$ where $n$ is the (integer) number of links in the slice. A pure quantum
state $\left|n\right\rangle $ is then associated with each spatial slice of
volume $na$, with $a$ fixed. These states are assumed to form an orthogonal
basis for the Hilbert space $\mathcal{T}$ of states. Locality implies that
the Hamiltonian acts only by coupling adjacent links, which are created or
destroyed by creation or annihiliation operators.

String bit models furnish an alternative to the discretization of path
integral formulations of a class of 2d quantum gravity models referred to as
Lorentzian gravity \cite{al, dgk}, and which are closely related to random
triangulation models. It has recently been demonstrated that the continuum
limits of these Lorentzian models can be obtained from these string bit
Hamiltonian models.

In this paper we extend this work to include string bit models of quantum
gravity coupled to matter. Specifically, we couple particles to gravity by
introducing new creation and annihilation operators that act to create and
destroy particles that reside on the links. In other words, we represent the
particle as a `coloured link', i.e., a link whose creation/annihilation
operator is distinguished from that of an empty string bit. We have here a
simple model of quantum gravity in which the energy of the particle(s)
influences the evolution of the spatial slice and vice versa. A given link
can hold more than one particle, and the particles can migrate from link to
link as the (string bit) universe expands/contracts (ie. loses or gains
links). The system is entirely quantum-mechanical, in that the particle can
be in an arbitrary superposition of locations on a collection of string bits.

We will start with a one-particle model in a closed universe in section~\ref%
{s2}. Then in section~\ref{s3} we will study an open-universe model with one
particle which moves in such a way that matter is homogeneously distributed
throughout the equal-time slice. Remarkably the one-particle closed and open
universe models behave like empty open and closed universes, respectively.
We will then turn our attention to a model with two distinguishable
particles in a closed universe in section~\ref{s4}. Again the particles give
rise to a homogeneous distribution. This turns out to behave like an empty
closed universe. In section~\ref{s5}, we will look into a closed-universe
model with indistinguishable bosons. We find that all the previous models
(as well as the empty universe models in ref.\cite{0108149}) are special
cases of the \emph{fractional sector model}, which thus plays the role of a
metamodel. We obtain a solution to the fractional sector model at the
continuum limit in section~\ref{s6}. Its transition amplitude turns out to
be exactly the same as the $sl(2)$ gravity model in ref.\cite{0108149}.  
Finally, we recapitulate our main results in section~\ref{s7}.

Our formalism is the same as that of ref.\cite{0108149}, and we will largely
follow the notation in that article unless otherwise specified. In
particular, we will always take the limit $N\rightarrow\infty$, and $a$ and $%
a^{\dagger}$ will always be operators annihilating and creating a link
representing an empty segment of an equal-time slice of a universe, or an
empty link in short, respectively. Two operators annihilating or creating
different kinds of links commute with each other unless both links represent
fermions, in which case the operators anti-commute.

\bigskip

\section{One Particle in a Closed Universe}

\label{s2}

A drifting particle in a closed universe is perhaps the simplest quantum
gravity model with matter. Consider an equal-time slice with a particle and $%
n$ empty links, where $n$ is non-negative. Its quantum state takes the form 
\[
|n+1\rangle _{c}:=\frac{1}{N^{(n+1)/2}}\mathrm{Tr}\,b^{\dagger}
(a^{\dagger})^{n}|\Omega\rangle , 
\]
where $b^{\dagger}$ creates a link representing the particle\footnote{%
Note that this $|n\rangle _{c}$ is different from the one in Ref.\cite%
{0108149}.}. $b$ and $b^{\dagger }$ satisfy the usual canonical commutation
relations. Furthermore, $b$ and $b^{\dagger}$ commute with $a$ and $%
a^{\dagger }$. (For more details, see Appendix~A.) The norm of $|n\rangle
_{c}$ is $1$: 
\begin{eqnarray}
\lim_{N \rightarrow \infty} \langle n | n \rangle_c = 1.  \label{2.0.1}
\end{eqnarray}

Our choice of the Hamiltonian is 
\begin{equation}
H:=H_{0}+\lambda H_{-1}+\lambda H_{1},  \label{2.1}
\end{equation}%
where 
\begin{eqnarray}
H_{0}:= &&\mathrm{Tr}\,a^{\dagger }a+V\mathrm{Tr}\,b^{\dagger }b,
\label{2.2} \\
H_{-1}:= &&\frac{1}{\sqrt{N}}\mathrm{Tr}\,(a^{\dagger })^{2}a+\frac{\xi }{%
\sqrt{N}}\mathrm{Tr}\,b^{\dagger }a^{\dagger }b+\frac{\xi }{\sqrt{N}}\mathrm{%
Tr}\,a^{\dagger }b^{\dagger }b,  \label{2.3}
\end{eqnarray}%
and 
\begin{equation}
H_{1}:=\frac{1}{\sqrt{N}}\mathrm{Tr}\,a^{\dagger }a^{2}+\frac{\xi }{\sqrt{N}}%
\mathrm{Tr}\,b^{\dagger }ab+\frac{\xi }{\sqrt{N}}\mathrm{Tr}\,b^{\dagger }ba,
\label{2.4}
\end{equation}%
and whose physical interpretation is as follows. The terms in $H_{0}$
measure the overall energy of a given closed universe with a single
particle, where the constant coefficient $V$ in eq.(\ref{2.2}) can be
conceived of as the ``mass'' of the particle. $H_{\pm 1}$ contains all terms
that create/destroy a link; the constant coefficient $\xi $ in eqs.(\ref{2.3}%
) and (\ref{2.4}) measures the relative ease by which an empty link adjacent
to the particle may be annihilated or created.

The subscripts 0, -1, and 1 in eq.(\ref{2.1}) are motivated by the following
observation. Since 
\begin{eqnarray*}
H_{0}|n\rangle & = & (n+V-1)|n\rangle_{c}, \\
H_{-1}|n\rangle & = & (n+2\xi -1 )|n+1\rangle _{c}
\end{eqnarray*}
for $n>0$, 
\[
H_{1}|1\rangle =0, 
\]
and 
\[
H_{1}|n\rangle =(n+2\xi -2)|n-1\rangle _{c} 
\]
for $n>1$, $H_{0}$, $H_{-1}$, and $H_{1}$ satisfy the commutation relations 
\[
\left[ H_{0},H_{-1}\right] =H_{-1}\;\mbox{and}\;\left[ H_{0},H_{1}\right]
=-H_{1} 
\]
for any value of $\xi $ and the relation 
\[
\left[ H_{1},H_{-1}\right] |n\rangle _{c}=\left\{ 
\begin{array}{ll}
2(H_{0}+2\xi -\frac{1}{2}-V)|n\rangle _{c} & \mbox{if $n \neq
   1$, or} \\ 
4\xi ^{2}|0\rangle & \mbox{if $n = 1$}.%
\end{array}
\right. 
\]
In particular, if $\xi =1/2$, then 
\[
\left[ H_{1},H_{-1}\right] =2(H_{0}+\frac{1}{2}-V). 
\]
We will make this simplifying assumption in the remainder of this section
and return to the general case in section~\ref{s6} after developing more
powerful mathematical machinery. Then $H_{0}+1/2-V$, $H_{-1}$, and $H_{1}$
may be thought of as the Virasoro generators $L_{0}$, $L_{-1}$, and $L_{1}$,
respectively. The state $|n\rangle _{c}$ corresponds to $|n\rangle _{\frac{1%
}{2}}$ in ref.\cite{0108149}, and the set of all $|n\rangle _{c}$'s span a
representation of $sl(2)$ with a highest weight $h=1/2$.

The transition amplitude $\tilde{G}(L,L^{\prime };T)$ at the continuum limit
defined as 
\[
\tilde{G}(L,L^{\prime };T)=\lim_{a\rightarrow 0}a\langle \frac{L^{\prime }}{a%
}|e^{-\frac{2TH}{a}}|\frac{L}{a}\rangle 
\]
has been shown to exist only if 
\begin{equation}
V=2\xi -\frac{1}{2},  \label{2.5}
\end{equation}
which is $1/2$ if $\xi =1/2$ \cite{0108149}. Then 
\begin{equation}
\tilde{G}(L,L^{\prime };T)=\frac{\sqrt{\Lambda }}{\sinh (\sqrt{\Lambda }T)}%
e^{-\sqrt{\Lambda }(L+L^{\prime })\coth (\sqrt{\Lambda }T)}I_{0}\left( \frac{%
2\sqrt{\Lambda LL^{\prime }}}{\sinh (\sqrt{\Lambda }T)}\right) ,  \label{2.6}
\end{equation}
where $\Lambda$ is the renormalized cosmological constant and is related to $%
\lambda$ by the formula 
\begin{equation}
\lambda := - \frac{1}{2} e^{- \frac{\Lambda}{2} a^2}. \label{2.7} 
\end{equation}
The emergence of the constraint (\ref{2.5}) suggests that the mass of a
particle is necessarily quantized once we quantize a space-time. Even more
unexpected is the result that there is only one legitimate value $V=1/2$
that the mass may take. Since the transition amplitude of an empty open
universe is identical to the expression given in eq.(\ref{2.6}), adding a
particle to an empty closed universe effectively turns it to an empty open
universe.

It is actually possible to solve this model at the continuum limit for an
arbitrary value of $\xi$; indeed, we will show in section~\ref{s6} that 
eq.(\ref{2.6}) holds true no matter what the value of $\xi$ is, but 
eq.(\ref{2.5}) is always a necessary condition for the continuum limit to be 
well defined.

\section{One Particle in an Open Universe}

\label{s3}

Let us turn our attention to an open universe instead of a closed one. The
quantum state of a typical equal-time slice now takes the form 
\[
|m,n\rangle _{o}:=\frac{1}{N^{(m+n+2)/2}}\bar{q}^{\dagger }(a^{\dagger
})^{m}b^{\dagger }(a^{\dagger })^{n}q^{\dagger }|\Omega \rangle , 
\]%
where $m$ and $n$ are arbitrary non-negative integers. This slice consists
of a particle with $m$ empty links on its left and $n$ empty ones on its
right. At the far left and right ends are the boundary links, whose creation
operators are denoted by $\bar{q}^{\dagger}$ and $q^{\dagger}$,
respectively. The norm of $|m,n\rangle _{o}$ is 1. Let $\mathcal{T}_{1}$ be
the Hilbert space spanned by all quantum states of the form 
\[
|n\rangle _{1}:=\sum_{k=0}^{n-1}|k,n-1-k\rangle _{o}, 
\]
where $n$ is an arbitrary positive integer. Its norm is given by the formula 
\[
\lim_{N\rightarrow \infty }\langle n|n\rangle _{1}=n. 
\]%
Physically speaking, $|n\rangle _{1}$ is a quantum state in which there are $%
n-1$ empty links and the particle may appear anywhere on the slice with
equal probability. In other words, the mass is homogeneously distributed
throughout the universe via this quantum superposition. We will see shortly
that $\mathcal{T}_{1}$ is indeed isomorphic to a representation space of $%
sl(2)$ with a highest weight 1.

Consider a Hamiltonian for this universe, which can again be written as 
\[
H:=H_{0}+\lambda H_{-1}+\lambda H_{1}, 
\]%
where, generalizing the closed universe case, we take%
\begin{eqnarray}
\lefteqn{H_{0}:=\mathrm{Tr}a^{\dagger }a-\frac{1}{4}\bar{q}^{\dagger }\bar{q}%
-\frac{1}{4}(q^{\dagger })^{t}q^{t}+V\mathrm{Tr}b^{\dagger }b}  \nonumber \\
&&-K\left\{ \frac{1}{N}\mathrm{Tr}\,a^{\dagger }b^{\dagger }ab+\frac{1}{N}%
\mathrm{Tr}\,b^{\dagger }a^{\dagger }ba+\frac{1}{N}\bar{q}^{\dagger
}b^{\dagger }b\bar{q}+\frac{1}{N}(q^{\dagger })^{t}(b^{\dagger
})^{t}b^{t}q^{t}\right\} ,  \label{3.1}
\end{eqnarray}%
\begin{eqnarray}
\lefteqn{H_{-1}:=\frac{1}{\sqrt{N}}\mathrm{Tr}(a^{\dagger })^{2}a+\frac{\xi 
}{\sqrt{N}}(\mathrm{Tr}b^{\dagger }a^{\dagger }b+\mathrm{Tr}a^{\dagger
}b^{\dagger }b)}  \nonumber \\
&&+\frac{\eta }{N\sqrt{N}}\left\{ \bar{q}^{\dagger }b^{\dagger }a^{\dagger }b%
\bar{q}+(q^{\dagger })^{t}(b^{\dagger })^{t}(a^{\dagger
})^{t}b^{t}q^{t}\right\} ,  \label{3.2}
\end{eqnarray}%
and 
\begin{eqnarray}
H_{1} & := & \frac{1}{\sqrt{N}} \mathrm{Tr} \, a^{\dagger} a^{2} + \frac{\xi%
}{\sqrt{N}} \left( \mathrm{Tr} \, b^{\dagger} ab + \mathrm{Tr}\,b^{\dagger
}ba\right)  \nonumber \\
&&+\frac{\eta }{N\sqrt{N}}\left\{ \bar{q}^{\dagger }b^{\dagger }ab\bar{q}%
+(q^{\dagger })^{t}(b^{\dagger })^{t}a^{t}b^{t}q^{t}\right\} .  \label{3.3}
\end{eqnarray}%
In eq.(\ref{3.1}), the first four terms in $H_{0}$ measure the total energy
of the open universe, its endpoints, and the single particle, again taken to
have mass $V$. The next set of terms measure the energy of the particle as
it moves between adjacent links, corresponding to the kinetic energy of the
particle as it moves around in the universe (its hopping motion). These
terms, proportional to $K$, constitute the usual Hubbard model satisfying
some open boundary conditions except that the particle is a boson instead of
a fermion \cite{9806019}, with $K$ the Hubbard constant. (Note that in the
model of section~\ref{s2}, the Hubbard-type terms are almost the same as the
identity operator. Indeed, 
\[
\frac{1}{\sqrt{N}}\mathrm{Tr}\,b^{\dagger }a^{\dagger }ba|n\rangle _{c}= 
\mathrm{Tr}\,b^{\dagger }b|n\rangle _{c} 
\]
if $n\neq 0$. Hence we may absorb these terms in $V\mathrm{Tr}\,b^{\dagger
}b $.) The superscript $t$ in eqs.(\ref{3.1}) to (\ref{3.3}) denotes the
transpose. For instance, 
\[
(q^{\dagger })^{t}(b^{\dagger })^{t}b^{t}q^{t}=q_{\mu _{1}}^{\dagger }b_{\mu
_{2}}^{\dagger \mu _{1}}b_{\mu _{3}}^{\mu _{2}}q^{\mu _{3}} 
\]%
in eq.(\ref{3.1}). As in the model in section~\ref{s2}, the terms $H_{\pm 1}$
destroy/create a single link, with $\xi $ measuring the relative ease that
an empty link adjacent to the particle is annihilated or created. Note the
presence of boundary terms in all of $H_{0}$, $H_{-1}$, and $H_{1}$.

It follows from eqs.(\ref{3.1}) and (\ref{3.2}) that 
\begin{equation}
H_{0}|n\rangle _{1}=(n-\frac{3}{2}+V-2K)|n\rangle _{1}  \label{3.4}
\end{equation}
and 
\begin{equation}
H_{-1}|n\rangle _{1}=(n+2\xi -2)|n+1\rangle _{1}+(\eta -\xi +1)\left(
|0,n+1\rangle _{o}+|n+1,0\rangle _{o}\right)  \label{3.5}
\end{equation}%
for $n>0$. In particular, $\mathcal{T}_{1}$ is invariant under the action of 
$H_{-1}$ if 
\begin{equation}
\eta =\xi -1.  \label{3.6}
\end{equation}
Then 
\[
H_{1}|1\rangle _{1}=0 
\]
and 
\[
H_{1}|n\rangle _{1}=(n+2\xi -2)|n-1\rangle _{1}+(\xi -1)\left( |0,n-1\rangle
_{o}+|n-1,0\rangle _{o}\right) 
\]%
for $n>1$ if we impose condition~(\ref{3.6}). Hence $\mathcal{T}_{1}$ is
invariant under the actions of both $H_{-1}$ and $H_{1}$ if 
\[
\xi =1\;\mbox{and}\;\eta =0. 
\]
In this case, $H_{0}+2K-V+1/2$, $H_{-1}$, and $H_{1}$ form a Lie algebra
isomorphic to $sl(2)$ if we treat $\mathcal{T}_{1}$ as a representation
space with a highest weight 1. It then follows from the calculations in ref. %
\cite{0108149} that the normalised transition amplitude 
\begin{eqnarray}
\tilde{G}(L,L^{\prime };T)=\lim_{a\rightarrow 0}\frac{a^{2}}{\sqrt{%
LL^{\prime }}}\langle \frac{L^{\prime }}{a}|e^{-\frac{2TH}{a}}|\frac{L}{a}
\rangle  \label{3.8}
\end{eqnarray}
exists if 
\begin{equation}
V-2K=\frac{3}{2},  \label{3.9}
\end{equation}
in which case 
\begin{equation}
\tilde{G}(L,L^{\prime };T)=\frac{\sqrt{\Lambda }}{\sinh (\sqrt{\Lambda }T)}%
e^{-\sqrt{\Lambda }(L+L^{\prime })\coth (\sqrt{\Lambda }T)}I_{1}\left( \frac{%
2\sqrt{\Lambda LL^{\prime }}}{\sinh (\sqrt{\Lambda }T)}\right) .
\label{1pclosed}
\end{equation}%
At the continuum limit, this model behaves exactly like an empty closed
universe.

Eq.(\ref{3.9}) implies that the mobility $|K|$ of the particle is related to 
its mass $V$. If $V > 3/2$, then the higher the mass, the higher the mobility.
If $V = 3/2$, then $K = 0$ and the particle cannot drift about the
equal-time slice. As it is reasonable to expect that no particle can have a
negative mass, eq.(\ref{3.9}) suggests that the value of $K$ has a lower bound 
of -3/4.

\section{Two Distinguishable Particles in a Closed Universe}

\label{s4}

Two-particle models display a rich variety of topologies and particle
statistics; the universe may be open or closed, the particles may be bosons
or fermions, and they may or may not be distinguishable.

Let us start with the model of two distinguishable particles in a closed
universe. We will show that a sector of this model could be identified with
the $sl(2)$ quantum gravity model. A typical quantum state of an equal-time
slice has the form 
\[
|m,n\rangle _{c,d}:=\frac{1}{N^{(m+n+2)/2}}\mathrm{Tr}\,b_{1}^{\dagger
}(a^{\dagger })^{m}b_{2}^{\dagger }(a^{\dagger })^{n}|\Omega \rangle , 
\]
where $m$ and $n$ are any non-negative integers, and $b_{1}^{\dagger }$ and $%
b_{2}^{\dagger }$ are the bosonic or fermionic creation operators of the two
particles, or the form 
\[
|-1,n\rangle _{c,d}:=\frac{1}{N^{(n+1)/2}}\mathrm{Tr}\,b_{12}^{\dagger
}(a^{\dagger })^{n}|\Omega \rangle , 
\]
where $n$ is again a non-negative integer, and $b_{12}^{\dagger }$ is the
creation operator of a link with both particles there. The norm of $%
|m,n\rangle _{c,d}$, where $m\geq -1$, is 1: 
\[
\lim_{N\rightarrow \infty }\langle m,n|m,n\rangle _{c,d}=1. 
\]
Let $\mathcal{T}_{1}$ be the Hilbert space spanned by all quantum states of
the form 
\[
|n\rangle _{1}:=\sum_{k=-1}^{n-2}|k,n-2-k\rangle _{c,d}, 
\]
where $n$ is an arbitrary positive integer. Its norm is $n$. Physically
speaking, this state represents a universe with $n$ links, one or two of
which may be link(s) with particle(s), and in which the two particles may
appear anywhere on the equal-time slice with equal probability. Thus this is
a model in which matter is homogeneously distributed, like the one in
section~\ref{s3}. We will justify the notations $\mathcal{T}_{1}$ and $%
|n\rangle _{1}$ shortly.

We will choose a Hamiltonian which is a variant of the Hubbard model. As
usual, 
\[
H:=H_{0}+\lambda H_{-1}+\lambda H_{1}. 
\]
$H_{0}$ is a collection of terms that leave the number of links unchanged: 
\begin{eqnarray}
\lefteqn{H_{0}:=\mathrm{Tr}a^{\dagger }a+V_{1}\mathrm{Tr}b_{1}^{\dagger
}b_{1}+V_{2}\mathrm{Tr}b_{2}^{\dagger }b_{2}+V_{12}\mathrm{Tr}
b_{12}^{\dagger }b_{12}}  \nonumber \\
&&-\frac{K_{1}}{N}\left( \mathrm{Tr}\,a^{\dagger }b_{1}^{\dagger }ab_{1}+ 
\mathrm{Tr}\,b_{1}^{\dagger }a^{\dagger }b_{1}a\right) -\frac{K_{2}}{N}
\left( \mathrm{Tr}\,a^{\dagger }b_{2}^{\dagger }ab_{2}+\mathrm{Tr}
\,b_{2}^{\dagger }a^{\dagger }b_{2}a\right)  \nonumber \\
&&-\frac{K_{12}}{N}\left( \mathrm{Tr}\,b_{1}^{\dagger }b_{2}^{\dagger
}ab_{12}+\mathrm{Tr}\,b_{2}^{\dagger }b_{1}^{\dagger }ab_{12}+\mathrm{Tr}
\,b_{1}^{\dagger }b_{2}^{\dagger }b_{12}a+\mathrm{Tr}\,b_{2}^{\dagger
}b_{1}^{\dagger }b_{12}a\right)  \nonumber \\
&&-\frac{K_{12}}{N}\left( \mathrm{Tr}\,a^{\dagger }b_{12}^{\dagger
}b_{1}b_{2}+\mathrm{Tr}\,b_{12}^{\dagger }a^{\dagger }b_{1}b_{2}+\mathrm{Tr}
\,a^{\dagger }b_{12}^{\dagger }b_{2}b_{1}+\mathrm{Tr}\,b_{12}^{\dagger
}a^{\dagger }b_{2}b_{1}\right) .  \label{4.2}
\end{eqnarray}
The first term on the R.H.S. of eq.(\ref{4.2}) is the volume energy of the
empty links. The next three terms are the total potential energy of the two
particles. The contact potential between the two particles is $V_{12} - V_1
- V_2$. The remaining terms describe the hopping motion of the two
particles. In particular, the terms proportional to $K_{1}$ describe the
movement of the first particle in empty space, those proportional to $K_{2}$
describe the movement of the second particle, and those proportional to $%
K_{12}$ describe how the two particles separate or come together.

In general, this $H_{0}$ does not leave $\mathcal{T}_{1}$ invariant.
Nevertheless, if the coefficients $V_{1}$, $V_{2}$, $V_{12}$, $K_{1}$, $K_{2}
$, and $K_{12}$ in $H_{0}$ satisfy a number of relations, $H_{0}$ will leave 
$\mathcal{T}_{1}$ invariant. Indeed, since 
\[
H_{0}|2\rangle _{1}=(V_{1}+V_{2}-4K_{12})|2\rangle
_{1}+(V_{12}+1-V_{1}-V_{2})|-1,1\rangle _{c,d},
\]
we get the relation 
\begin{equation}
V_{12}=V_{1}+V_{2}-1.  \label{4.3}
\end{equation}%
This relation in turn yields 
\[
H_{0}|3\rangle _{1}=(V_{1}+V_{2}-2K_{1}-2K_{2}+1)|3\rangle
_{1}+(K_{1}+K_{2}-2K_{12})\left( |-1,2\rangle _{c,d}+|3\rangle _{1}\right) ,
\]
which further implies 
\begin{equation}
K_{12}=\frac{1}{2}(K_{1}+K_{2}).  \label{4.4}
\end{equation}%
Assuming both eqs.(\ref{4.3}) and (\ref{4.4}) then yields 
\[
H_{0}|1\rangle _{1}=(V_{1}+V_{2}-1)|1\rangle _{1}
\]
and 
\[
H_{0}|n\rangle _{1}=(n+V_{1}+V_{2}-2K_{1}-2K_{2}-2)|n\rangle _{1}
\]
for $n>1$. Then $\mathcal{T}_{1}$ is manifestly invariant under the action
of $H_{0}$.  Henceforth we will assume the validity of eqs.(\ref{4.3}) and
(\ref{4.4}).

A reasonable choice of $H_{-1}$ which respects locality in action and
translational invariance is 
\begin{eqnarray}
\lefteqn{H_{-1} := N^{-\frac{1}{2}} \{ \mathrm{Tr} \, (a^{\dagger })^2 a +
\x_1 ( \mathrm{Tr} a^{\dagger} b_1^{\dagger} b_1 + \mathrm{Tr} b_1^{\dagger}
a^{\dagger} b_1 )}  \nonumber \\
& & + \xi _2 (\mathrm{Tr} \, a^{\dagger} b_2^{\dagger} b_2 + \mathrm{Tr} \,
b_2^{\dagger} a^{\dagger} b_2) + \xi _{12} (\mathrm{Tr} \, a^{\dagger}
b_{12}^{\dagger} b_{12} + \mathrm{Tr} \, b_{12}^{\dagger} a^{\dagger} b_{12})
\nonumber \\
& & + \left. \eta (\mathrm{Tr} \, b_1^{\dagger} b_2^{\dagger} b_{12} + 
\mathrm{Tr} \, b_2^{\dagger} b_1^{\dagger} b_{12}) \right\}  \label{4.6}
\end{eqnarray}
and this corresponds to the creation of a single link in a universe with two
particles. $\xi_1$, $\xi_2$, $\xi_{12}$, and $\eta$ in eq.(\ref{4.6}) are
constants measuring the relative ease an empty link may be created from a
link with particle(s). Then 
\begin{eqnarray*}
H_{-1} | n \rangle_1 & = & (n + 2 \xi_1 + 2 \xi_2 - 3) | n+1 \rangle_1 + (2
\xi_{12} - 2 \xi_1 - 2 \xi_2 + 2) |-1, n \rangle_{c,d} \\
& & + (\eta - \xi_1 - \xi_2 + 1) \left( |0, n-1 \rangle_{c,d} + |n-1, 0
\rangle_{c,d} \right).
\end{eqnarray*}
Hence $\mathcal{T}_1$ is invariant under the action of $H_{-1}$ if and only
if 
\begin{eqnarray}
\eta = \xi_{12} = \xi_1 + \xi_2 - 1.  \label{4.7}
\end{eqnarray}
$H_1$ is the Hermitian conjugate of $H_{-1}$ and hence has the form 
\begin{eqnarray*}
\lefteqn{H_1 := N^{-\frac{1}{2}} \le\{ \Tr a^{\da} a^2 + \x_1 ( \Tr
b_1^{\da} a b_1 + \Tr b_1^{\da} b_1 a ) \ri. } \\
& & + \xi_2 ( \mathrm{Tr} \, b_2^{\dagger} a b_2 + \mathrm{Tr} \,
b_2^{\dagger} b_2 a ) + \xi_{12} ( \mathrm{Tr} \, b_{12}^{\dagger} a b_{12}
+ \mathrm{Tr} \, b_{12}^{\dagger} b_{12} a ) \\
& & + \left. \eta (\mathrm{Tr} \, b_{12}^{\dagger} b_1 b_2 + \mathrm{Tr} \,
b_{12}^{\dagger} b_2 b_1) \right\}.
\end{eqnarray*}
Its action on $\mathcal{T}_1$ reads 
\[
H_1 | 1 \rangle_1 = 0 
\]
and 
\[
H_1 | n \rangle_1 = (n + 2 \xi_1 + 2 \xi_2 - 3) | n-1 \rangle_1 + (2 \eta +
2 \xi_{12} - 2 \xi_1 - 2 \xi_2 + 1) |-1, n-2 \rangle_{c,d} 
\]
for $n \geq 2$. Thus $\mathcal{T}_1$ is left invariant if 
\begin{eqnarray}
2 \eta + 2 \xi_{12} = 2 \xi_1 + 2 \xi_2 - 1.  \label{4.8}
\end{eqnarray}
Combining eqs.(\ref{4.7}) and (\ref{4.8}) yields 
\begin{eqnarray}
\xi_1 + \xi_2 = \frac{3}{2}, \; \mbox{and} \; \xi_{12} = \eta = \frac{1}{2}.
\label{4.9}
\end{eqnarray}

Under the constraints (\ref{4.3}), (\ref{4.4}), and (\ref{4.9}), 
\bea
   H_{-1} |n\rangle_1 & = & n |n+1\rangle_1, \\
   H_1 |n\rangle_1 & = & n |n-1\rangle_1, \\
   \le( H_0 + 2 - V_1 - V_2 \ri) |1\rangle_1 & = & |1\rangle_1, \; 
   \mbox{and} \\
   \le( H_0 + 2 - V_1 - V_2 \ri) |n\rangle_1 & = & (n - 2 K_1 - 2 K_2) 
   |n\rangle_1 
\eea
if $n > 1$.  Then $H_{0}+2-V_{1}-V_{2}$, $H_{-1}$, and $H_{1}$ form a Lie 
algebra isomorphic to $sl(2)$ provided that
\beq
K_{1}+K_{2}=0.
\la{4.4a}
\eeq
In addition, $\mathcal{T}_{1}$ is a unitary representation with a highest 
weight~1. Hence the normalised transition amplitude at the continuum limit, 
which may also be defined as in eq.(\ref{3.8}), is again 
\begin{equation}
\tilde{G}(L,L^{\prime };T)=\frac{\sqrt{\Lambda }}{\sinh (\sqrt{\Lambda }T)}%
e^{-\sqrt{\Lambda }(L+L^{\prime })\coth (\sqrt{\Lambda }T)}I_{1}\left( \frac{%
2\sqrt{\Lambda LL^{\prime }}}{\sinh (\sqrt{\Lambda }T)}\right) .
\label{2pdclosed}
\end{equation}
provided that 
\beq
V_{1}+V_{2}-2K_{1}-2K_{2}=2.
\la{4.10}
\eeq
Note that even if eq.(\ref{4.4a}) does not hold true, the transition
amplitude at the continuum limit is still given by eq.(\ref{2pdclosed})
provided that the constraint (\ref{4.10}) is valid. We will show this in
section~\ref{s6} when we consider the fractional sector model.

\section{Two Indistinguishable Bosons in a Closed Universe}

\label{s5}

Let us turn our attention to the case in which the two particles are
indistinguishable bosons. This situation is qualitatively distinct from
previous models (which are all isomorphic to an $sl(2)$ gravity model if the
matter is homogeneously distributed) and solving this model will require new
techniques. Instead of obtaining an energy spectrum first and then passing
to the continuum limit, we will look for the eigenenergies and eigenstates
at the continuum limit directly.

Consider an equal-time slice with two indistinguishable bosons. If they lie
on different links, and one of the bosons is separated from the other by $m$
empty links on one side and $n$ empty links on the other, its quantum state
is 
\[
|m,n\rangle _{c,i}:=\frac{1}{N^{(m+n+2)/2}}\mathrm{Tr} \,b_{1}^{\dagger
}(a^{\dagger })^{m}b_{1}^{\dagger }(a^{\dagger })^{n} |\Omega \rangle , 
\]
where $b_{1}^{\dagger }$ creates a link with a boson, and $m$ and $n$ are
non-negative integers. On the other hand, if they lie on the same link and
there are $n$ empty links in the equal-time slice, then its quantum state is 
\[
|-1,n\rangle _{c,i}:=\frac{1}{N^{(n+1)/2}}\mathrm{Tr}\,b_{2}^{\dagger}
(a^{\dagger })^{n} |\Omega \rangle , 
\]
where $b_{2}^{\dagger}$ creates a link with two bosons, and $n$ is a
non-negative integer. Note that the state $|0,n\rangle _{c,i}$ is distinct
from the state $|-1,n\rangle _{c,i}$: the former consists of $n+2$ links,
and there are two adjacent links with a single particle on each, whereas the
latter consists of $n+1$ links with two particles on the same link. The
norms of these states are 1 or 2: 
\[
\lim_{N\rightarrow \infty }\langle m,n|m,n\rangle _{c,i}=\left\{ 
\begin{array}{ll}
1 & \mbox{if $m \neq n$ or} \\ 
2 & \mbox{if $m = n$}%
\end{array}
\right. 
\]

A quantum state with a homogeneous matter distribution takes either the form 
\begin{equation}
|n+1\rangle _{1^{\prime }}:=\frac{1}{2}|-1,2n\rangle
_{c,i}+\sum_{k=0}^{n-1}|k,2n-1-k\rangle _{c,i}  \label{5.1}
\end{equation}
if the number of links $2n+1$ is odd, or 
\begin{equation}
\frac{1}{2}|-1,2n+1\rangle _{c,i}+\sum_{k=0}^{n-1}|k,2n-k\rangle _{c,i}+ 
\frac{1}{2}|n,n\rangle _{c,i}  \label{5.1a}
\end{equation}
if the number of links $2n+2$ is even. The two bosons may appear anywhere in
the equal-time slice with equal probability. The norm of $|n\rangle
_{1^{\prime }}$ is given by the formula 
\begin{eqnarray}
\lim_{N\rightarrow \infty }\langle n|n\rangle _{1^{\prime }}=n+\frac{1}{2}.
\label{5.1b}
\end{eqnarray}

We define $\mathcal{T}_{1^{\prime }}$ to be the Hilbert space spanned by all
states of the form (\ref{5.1}) with $n\geq 0$. We will see that while there
is a non-trivial Hamiltonian that leaves the space $\mathcal{T}_{1^{\prime
}} $ invariant, the various terms of the Hamiltonian do not form the Lie
algebra $sl(2)$. Consequently, $\mathcal{T}_{1^{\prime }}$ is different from 
$\mathcal{T}_{1}$, though they display some similar properties.

As before, there are three parts in the Hamiltonian 
\[
H=H_{0}+\lambda H_{-1}+\lambda H_{1}. 
\]
corresponding to an overall volume/particle energy, and the energy to
create/destroy links and/or particles. The simplest choice for $H_{0}$ that
has the fewest number of operators is 
\begin{eqnarray}
& H_{0} := \mathrm{Tr} a^{\dagger} a + V \mathrm{Tr} b_{1}^{\dagger} b_{1} +
V_{2} \mathrm{Tr} b_{2}^{\dagger} b_{2} - \frac{K}{N} \left( \mathrm{Tr}
a^{\dagger} b_{1}^{\dagger} ab_{1} + \mathrm{Tr} b_{1}^{\dagger} a^{\dagger}
b_{1} a \right) &  \nonumber \\
& - \frac{K_{2}}{N} \left( \mathrm{Tr} b_{1}^{\dagger} b_{1}^{\dagger}
ab_{2} + \mathrm{Tr} b_{1}^{\dagger} b_{1}^{\dagger} b_{2} a + \mathrm{Tr}
a^{\dagger} b_{2}^{\dagger} b_{1} b_{1} + \mathrm{Tr} b_{2}^{\dagger}
a^{\dagger} b_{1} b_{1} \right) . &  \label{5.2}
\end{eqnarray}
In this formula, $V$ is the mass of a boson, and $V_{2}-2V$ is the contact
potential between the two bosons. $K$ and $K_{2}$ are the analogues of
Hubbard constants that measure how easy a boson hops from one link to an
adjacent link or how easy the two bosons move apart or come together. Its
action on $\mathcal{T}_{1^{\prime }}$ is straightforwardly computed to be 
\begin{eqnarray}
H_{0}|1\rangle _{1^{\prime }} &=&V_{2}|1\rangle _{1^{\prime }},
\label{5.2.1} \\
H_{0}|2\rangle _{1^{\prime }} &=&\left( 1+2V-2K-K_{2}\right) |2\rangle
_{1^{\prime }}  \nonumber \\
&&+\left( \frac{1}{2}-V+\frac{1}{2}V_{2}+K-\frac{3}{2}K_{2}\right)
|-1,2\rangle _{c,i},  \label{5.3}
\end{eqnarray}
and 
\begin{eqnarray}
\lefteqn{H_{0}|n\rangle _{1\prime }=(-3+2n+2V-4K)|n\rangle _{1\prime }} 
\nonumber \\
&&+\left( \frac{1}{2}-V+\frac{V_{2}}{2}+4K-4K_{2}\right)
|-1,2n-2\rangle_{c,i}  \nonumber \\
&&+(2K-K_{2})|0,2n-1\rangle _{c,i}  \label{5.4}
\end{eqnarray}
for $n>2$. Then $\mathcal{T}_{1^{\prime }}$ is invariant if 
\begin{equation}
K_{2}=2K\;\;\mbox{and}\;\;V_{2}=4K+2V-1.  \label{5.5}
\end{equation}%
In particular, there is contact interaction between the two bosons. We will
henceforth impose the constraints (\ref{5.5}). Then eqs.(\ref{5.3}) and (\ref%
{5.4}) imply that 
\begin{equation}
H_{0}|n\rangle _{1^{\prime }}=(-3+2n+2V-4K)|n\rangle _{1^{\prime }}
\label{5.6}
\end{equation}
for $n>1$. From eq.(\ref{5.2.1}), we see that eq.(\ref{5.6}) holds true for $%
n=1$ only if $K=0$. However we will not impose this condition; instead we
will develop a more powerful method to show that one could still evaluate,
at the continuum limit, the transition amplitude of this model in which the
bosons may hop from one link to another.

The most general $H_{-1}$ which is local, translationally invariant and
exhibits left-right symmetry in action is 
\begin{eqnarray}
H_{-1} & := & \frac{1}{N} \mathrm{Tr} \, (a^{\dagger})^{3} a + \frac{\xi_{1}%
}{N} \left\{ \mathrm{Tr} \, (a^{\dagger})^{2} b_{1}^{\dagger} b_{1} + 
\mathrm{Tr} \, b_{1}^{\dagger} (a^{\dagger})^{2} b_{1} \right\} + \frac{%
\xi_{2}}{N} \mathrm{Tr} \, a^{\dagger}b_{1}^{\dagger} a^{\dagger} b_{1} 
\nonumber \\
& & + \frac{\xi_{3}}{N} \left\{ \mathrm{Tr} \, ( a^{\dagger})^{2}
b_{2}^{\dagger} b_{2} + \mathrm{Tr} \, b_{2}^{\dagger} (a^{\dagger})^{2}
b_{2} \right\} + \frac{\xi_{4}}{N} \mathrm{Tr} \, a^{\dagger}
b_{2}^{\dagger} a^{\dagger} b_{2}  \nonumber \\
& & + \frac{\eta_{1}}{N} \left\{ \mathrm{Tr} \, a^{\dagger}
(b_{1}^{\dagger})^{2} b_{2} + \mathrm{Tr} \, (b_{1}^{\dagger})^{2}
a^{\dagger} b_{2} \right\} + \frac{\eta_{2}}{N} \mathrm{Tr} \,
b_{1}^{\dagger} a^{\dagger} b_{1}^{\dagger} b_{2} .  \label{5.7}
\end{eqnarray}
Its net effect is to create two more links. The six constants $\xi _{1}$, $%
\xi _{2}$, $\xi _{3}$, $\xi _{4}$, $\eta _{1}$, and $\eta _{2}$ measure the
relative ease by which the various splitting processes take place. Its
action on $\mathcal{T}_{1^{\prime }}$ reads 
\[
H_{-1}|1\rangle _{1^{\prime }}=\left( \xi _{3}+\frac{1}{2}\xi _{4}\right)
|-1,2\rangle _{c,i}+\left( \eta _{1}+\frac{1}{2}\eta _{2}\right) |0,1\rangle
_{c,i} 
\]
and 
\begin{eqnarray*}
\lefteqn{H_{-1}|n\rangle _{1\prime }=(-5+2n+4\xi _{1}+2\xi _{2})|n\rangle
_{1\prime }} \\
&&+\left( \frac{3}{2}-2\xi _{1}-\xi _{2}+\xi _{3}+\frac{1}{2}\xi _{4}\right)
|-1,2n\rangle _{c,i} \\
&&+\left( 2-2\xi _{1}-2\xi _{2}+\eta _{1}\right) |0,2n-1\rangle
_{c,i}+\left( 1-2\xi _{1}+\frac{1}{2}\eta _{2}\right) |1,2n-2\rangle _{c,i}
\end{eqnarray*}%
for $n>1$. Hence $\mathcal{T}_{1^{\prime }}$ is left invariant if 
\begin{eqnarray}
2\xi _{3}+\xi _{4} &=&-3+4\xi _{1}+2\xi _{2},  \nonumber \\
\eta _{1} &=&-2+2\xi _{1}+2\xi _{2},  \nonumber \\
\eta _{2} &=&-2+4\xi _{1}.  \label{5.8}
\end{eqnarray}%
We will assume that eqs.(\ref{5.8}) hold true in all the following
calculation.

$H_{1}$ is the Hermitian conjugate of $H_{-1}$ defined in eq.(\ref{5.7}): 
\begin{eqnarray}
\lefteqn{H_{1} := \frac{1}{N} \mathrm{Tr} \, a^{\dagger} a^{3} + \frac{%
\xi_{1}}{N} \left\{ \mathrm{Tr} \, b_{1}^{\dagger} a^{2} b_{1} + \mathrm{Tr}
\, b_{1}^{\dagger} b_{1} a^{2} \right\} + \frac{\xi_{2}}{N} \mathrm{Tr} \,
b_{1}^{\dagger} a b_{1}a }  \nonumber \\
& & + \frac{\xi_{3}}{N} \left\{ \mathrm{Tr} \, b_{2}^{\dagger} a^{2} b_{2} + 
\mathrm{Tr} \, b_{2}^{\dagger} b_{2} a^{2} \right\} + \frac{\xi_{4}}{N} 
\mathrm{Tr} \, b_{2}^{\dagger} a b_{2} a  \nonumber \\
& & + \frac{\eta_{1}}{N} \left\{ \mathrm{Tr} \, b_{2}^{\dagger} a
(b_{1})^{2} + \mathrm{Tr} \, b_{2}^{\dagger} (b_{1})^{2} a \right\} + \frac{%
\eta_{2}}{N} \mathrm{Tr} \, b_{2}^{\dagger }b_{1}ab_{1} .  \label{5.9}
\end{eqnarray}
Then 
\[
H_{1}|1\rangle _{1^{\prime }}=0 
\]%
and 
\begin{eqnarray*}
\lefteqn{H_{1}|n\rangle _{1\prime }=(-5+2n+4\xi _{1}+2\xi _{2})|n-1\rangle
_{1\prime }} \\
&&+\left( \frac{1}{2}-2\xi _{1}-\xi _{2}+\xi _{3}+\frac{1}{2}\xi _{4}+2\eta
_{1}+\eta _{2}\right) |-1,2n-4\rangle _{c,i}
\end{eqnarray*}%
for $n>1$. Hence $\mathcal{T}_{1^{\prime }}$ is invariant under the action
of $H_{1}$ if 
\begin{eqnarray}
2\xi _{3}+\xi _{4}+4\eta _{1}+2\eta _{2}=-1+4\xi _{1}+2\xi _{2}.
\label{5.10}
\end{eqnarray}
Combining eqs.(\ref{5.8}) and (\ref{5.10}) yields 
\begin{eqnarray}
\xi _{2} &=&\frac{7}{4}-2\xi _{1},  \nonumber \\
2\xi _{3}+\xi _{4} &=&\frac{1}{2},  \nonumber \\
\eta _{1} &=&\frac{3}{2}-2\xi _{1},\;\mbox{and}  \nonumber \\
\eta _{2} &=&-2+4\xi _{1}.  \label{5.11}
\end{eqnarray}%
Under these constraints, 
\begin{eqnarray}
H_{-1}|n\rangle _{1^{\prime }} &=&\left( 2n-\frac{3}{2}\right) |n+1\rangle
_{1^{\prime }},  \nonumber \\
H_{1}|1\rangle _{1^{\prime }} &=&0,\;\mbox{and}  \nonumber \\
H_{1}|n\rangle _{1^{\prime }} &=&\left( 2n-\frac{3}{2}\right) |n-1\rangle
_{1^{\prime }}\;\mbox{for $n > 1$.}  \label{5.12}
\end{eqnarray}%
Eqs.(\ref{5.6}) and (\ref{5.12}) then imply that $H_{0}$ (or $H_{0}+4K-2V$), 
$H_{-1}$, and $H_{1}$ do not form a Lie algebra; a different method is
needed to find the transition amplitude at the continuum limit. We will
describe this method in the following section.

Consider next the set of states spanned by all states of the form (\ref{5.1a}%
) with $n\geq 0$. This consists of all states with an even number of links.
It is straightforward to show that the resulting Hilbert space is not
invariant under the action of $H_{0}$ unless the coupling constant $K$ (and
consequently $K_2$) vanishes, which clearly trivializes the system.

\section{The Fractional Sector Model}

\label{s6}

Consider again the model of one particle in a closed universe. Let 
\[
\phi =\sum_{n=0}^{\infty }a_{n+2\xi }|n+1\rangle , 
\]%
where the coefficients $a_{2\xi }$, $a_{1+2\xi }$, $a_{2+2\xi }$, \ldots ,
and so on are complex constants, be an eigenstate of the Hamiltonian of this
model. (We will explain the meaning of the subscript $n+2\xi $ shortly.) 
The quantum state $\phi $\ can be regarded as a superposition of closed 
universes of all possible string bit lengths, each universe containing a 
single particle.

The eigenequation 
\[
H\phi =E\phi , 
\]%
where $E$ is the eigenenergy, leads to 
\begin{eqnarray}
\lefteqn{\lambda (n+2\xi +1)a_{n+2\xi +2}}  \nonumber \\
&&+\left( n+2\xi +\frac{1}{2}+E^{\prime }-E\right) a_{n+2\xi +1}+\lambda
(n+2\xi )a_{n+2\xi }=0,  \label{6.0.1}
\end{eqnarray}
where 
\begin{eqnarray}
E^{\prime }=V-2\xi +\frac{1}{2}  \label{6.0.1.1}
\end{eqnarray}
and $n$ is any non-negative integer, and 
\begin{eqnarray}
\lambda 2\xi a_{2\xi +1}+\left( 2\xi -\frac{1}{2}+E^{\prime }-E\right)
a_{2\xi }=0.  \label{6.0.2}
\end{eqnarray}
Similarly, if 
\[
\phi =\sum_{n=0}^{\infty }a_{n+\frac{1}{4}}|n+1\rangle 
\]
(Again we will explain the subscript $n+1/4$ shortly) and $E$ are an
eigenstate and eigenenergy of the Hamiltonian of the model of two
indistinguishable particles in a closed universe, then the complex
coefficients $a_{1/4}$, $a_{5/4}$, $a_{9/4}$, \ldots , and so on satisfy the
difference equations 
\begin{equation}
\lambda \left( n+\frac{9}{4}\right) a_{n+\frac{9}{4}}+\left( n+\frac{5}{4}+
\frac{E^{\prime }}{2}-\frac{E}{2}\right) a_{n+\frac{5}{4}} + \lambda \left( 
n+\frac{1}{4}\right) a_{n+\frac{1}{4}}=0,  \label{6.0.3}
\end{equation}
where 
\begin{eqnarray}
\frac{E^{\prime}}{2}=V-2K-\frac{3}{4}  \label{6.0.5}
\end{eqnarray}
and $n$ is any non-negative integer, and 
\begin{equation}
\lambda \frac{5}{4}a_{\frac{5}{4}}+\left( \frac{1}{4} + \frac{E'}{2} - 
\frac{E}{2} - 4K \right) a_{\frac{1}{4}}=0.  \label{6.0.4}
\end{equation}

Eqs.(\ref{2.0.1}), (\ref{6.0.1}), (\ref{6.0.2}), (\ref{5.1b}), (\ref{6.0.3}), 
and (\ref{6.0.4}) show that the two models are just special cases of a
subsuming {\em fractional sector model} whose Hilbert space is spanned by an
orthogonal set of vectors of the form $|n\rangle_h$, where $n$ is an
arbitrary positive integer and $h$ is a fixed real number. The norms of
these vectors are given by 
\begin{eqnarray}
\langle n+1 | n+1 \rangle_h = \frac{\Gamma (n+2h)}{n! \Gamma (2h)}.
\label{6.0.6}
\end{eqnarray}
The eigenstates of the Hamiltonian take the form 
\begin{eqnarray}
\phi = \sum_{n=0}^{\infty} a_{n+f} |n+1\rangle_h,  \label{6.2}
\end{eqnarray}
where $0 \leq f < 1$ and the coefficients $a_f$, $a_{f+1}$, $a_{f+2}$,
\ldots, and so on are complex constants satisfying the difference equations 
\begin{eqnarray}
\lambda (n + f + 2h) a_{n+f+2} + (n + f + h + E^{\prime}- E) a_{n+f+1} +
\lambda n a_{n+f} = 0  \label{6.6.1}
\end{eqnarray}
for $n$ = 0, 1, 2, \ldots, and so on, and 
\begin{eqnarray}
\lambda (f + 2h - 1) a_{f+1} + (f + h - 1 + E^{\prime}- E + V_0) a_f = 0,
\label{6.7.1}
\end{eqnarray}
where $V_0$ is a real constant. $h = 1/2$, $f = 2 \xi$, and $V_0 = 0$ for
the model in section~\ref{s2}; $h = 1$, $f = 0$, $V_0 = 2 K_1 + 2 K_2$, and
$E' = V_1 + V_2 - 2 K_1 - 2 K_2 - 1$ for the model in section~\ref{s4}; and 
$h = 1$, $f = 1/4$, and $V_0 = -4K$ for the model in section~\ref{s5}. 
Furthermore, if $f = V_0 = 0$, then the fractional sector model reduces to 
the $sl(2)$ gravity model in Ref.\cite{0108149}.  (Though the norms of 
$|n\rangle_{1^{\prime}}$ as shown in eq.(\ref{5.1b}) do not agree with 
eq.(\ref{6.0.6}), the difference is negligible at the continuum limit.)

\subsection{A General Solution}

There is a general solution to eqs.(\ref{6.6.1}) and (\ref{6.7.1}) in terms of
a generating function.  A straightforward analysis of the large-$n$ behavior 
of $a_n$ reveals that 
\begin{equation}
a_{n} = b^{\prime}_n p^{n}+ b_n p^{-n},  \label{6.4}
\end{equation}
where 
\[
p = \frac{-1 + \sqrt{1 - 4 \lambda^2}}{2 \lambda} 
\]
and $b^{\prime}_n$ and $b_n$ grow with $n$, if at all, polynomially.
Normalizability of $\phi$ forces all $b^{\prime}_n$ to vanish. Let 
\begin{equation}
\psi := \sum_{n=0}^{\infty} b_{n+f} x^{n+f}.  \label{6.8}
\end{equation}
be a generating function of $b_n$ (and hence $a_n$).  Then eqs.(\ref{6.6.1}) 
and (\ref{6.7.1}) imply, after some algebra, that 
\begin{eqnarray}
& x ( \lambda x^2 + px + \lambda p^2 ) \frac{d\psi}{dx} + \left\lb 2 \lambda
p^2 (h-1) + p (h-1+E^{\prime}-E)x \right\rb \psi &  \nonumber \\
& = b_f p \left\lb -V_0 x^{f+1} + \lambda p (f + 2h - 2) x^f \right\rb. &
\label{6.16}
\end{eqnarray}
Multiplying both sides of this equation by an integrating factor yields
\begin{eqnarray}
   \lefteqn{\frac{d}{dx} \left\{ \psi \le\lb \frac{x^2}{(1-x)(p^2 - x)} 
   \ri\rb^{h-1} \left( \frac{1-x}{p^2 - x} \right)^{\frac{(E-E')p}
   {\lambda (p^2 - 1)}} \right\} = b_f p x^{f-1} } \nn \\
   & & \cdot \frac{\l p (f + 2h - 2) - V_0 x}{(1-x) (p^2 - x)}
   \le\lb \frac{x^2}{(1-x)(p^2 - x)} \ri\rb^{h-1} \left( \frac{1-x}
   {p^{2}-x} \right)^{\frac{(E-E')p}{\lambda (p^{2}-1)}}.  
\label{6.9}
\end{eqnarray}
Let
\[ q := \frac{(E-E')p}{\lambda (p^{2}-1)} - h. \]
Then we may use the binomial series 
\[ \left( 1-\frac{1-p^{2}}{1-x} \right)^{-(q+2h)} =
   \sum_{r=0}^{\infty} \frac{\Gamma (q+2h+r)}{r! \Gamma (q+2h)} 
   \left( \frac{1-p^{2}}{1-x} \right)^{r} \]
to rewrite eq.(\ref{6.9}) as
\beq
   \lefteqn{\frac{d}{dx} \left\{ \psi \le\lb \frac{x^2}{(1-x)(p^2 - x)} 
   \ri\rb^{h-1} \left( \frac{1-x}{p^2 - x} \right)^{q+h} \right\} } \nn \\
   & = & b_f \l p^2 (f + 2h - 2) \sum_{r=0}^{\ift} 
   \frac{\G (q+2h+r)}{r! \G (q+2h)} (1-p^2)^r \frac{x^{f+2h-3}}{(1-x)^{r+2h}}
   \nn \\
   & & - b_f V_0 p \sum_{r=0}^{\ift} 
   \frac{\G (q+2h+r)}{r! \G (q+2h)} (1-p^2)^r \frac{x^{f+2h-2}}{(1-x)^{r+2h}}.
\la{6.12}
\eeq
Since
\[ \int \frac{x^{f+h'-3} dx}{(1-x)^{r+h'}} =
   \frac{x^{f+h'-2}}{f+h'-2} \, _2 F_1 (f + h' - 2, r + h'; f + h' - 1; x) 
   + A, \]
where $h'$ is any positive number, $_{2}F_{1} ( a,b;c,x) $ is the 
hypergeometric function
\[ 
_{2}F_{1} (a,b;c;x) = \frac{\Gamma (c)}{\Gamma (b) \Gamma (a)} 
\sum_{n=0}^{\infty} \frac{\Gamma (b+n) \Gamma (a+n)}{\Gamma (c+n)}
\frac{x^{n}}{n!},  
\]
and $A$ is a constant of integration, we may integrate eq.(\ref{6.12}) to
\begin{eqnarray*}
   \lefteqn{ \psi = \frac{(1-x)^{q+2h-1}}{x^{2(h-1)} (p^2 - x)^{q+1}} 
   \le\{ A + \sum_{r=0}^{\ift} 
   \frac{b_f p \Gamma (q+2h+r) (1-p^2)^r}{r! \Gamma (q+2h)} \ri. } \\
   & & \cdot \le\lb \l p \, _2 F_1 (f+2h-2, r+2h; f+2h-1; x) x^{f+2h-2} 
   \ri. \\ 
   & & \le. \le. - V_0 \, _2 F_1 (f+2h-1, r+2h; f+2h; x) x^{f+2h-1} \ri\rb 
   \ri\}.
\end{eqnarray*}

\subsection{The Continuum Limit}

There is a solution to the fractional sector model at the continuum limit, in 
which only terms of the form $(an)^r$, where $r$ is an integer, survive in 
$b_n$. Thus we may approximate eq.(\ref{6.16}) by making a change of the 
independent variable from $x$ to $w = 1-x$ which is of order $a$.  Writing 
\begin{eqnarray}
E - E^{\prime}:= (h + R) \sqrt{\Lambda} a  \label{6.20}
\end{eqnarray}
for some parameter $R$, we then obtain 
\[
\left( \frac{1}{2} w^2 - \sqrt{\Lambda} a w \right) \frac{d\psi}{dw} +
\left\lb (1 - h) w - (R + 1) \sqrt{\Lambda} a \right\rb \psi \simeq \frac{1}{
2} (2h-1) c_0,
\]
where $\La$ was defined in eq.(\ref{2.7}) and
\begin{eqnarray}
c_0 := \frac{-2V - f - 2h + 2}{2h-1} b_f  \label{6.18}
\end{eqnarray}
if $h \neq 1/2$, which we will assume in the next paragraph.

The solution to this first-order ordinary differential equation is 
\begin{eqnarray}
\psi(w) & = & c_0 (2h-1) \sum_{r=0}^{\infty} \frac{\Gamma(R+1)}{r!
\Gamma(R-r+1)} \frac{(2 \sqrt{\Lambda} a)^r}{r + 2h - 1} \frac{(w - 2 \sqrt{
\Lambda} a)^{R-r}}{w^{R+1}}  \nonumber \\
& & + A \frac{(w - 2 \sqrt{\Lambda} a)^{R + 2h - 1}}{w^{R+1}},  
\label{6.17}
\end{eqnarray}
where $A$ is a constant of integration.  If $c_0 = 0$, then $b_f = a_f = 0$,
and it follows from eqs.(\ref{6.6.1}) and (\ref{6.7.1}) that $\ps = 0$.  
Consequently, $A = 0$.  Thus we may use the identity\footnote{The L.H.S. of 
this identity should read 
\[
v^{1-2h} \lim_{\epsilon \rightarrow 0^+} \left\{ \int_{\epsilon}^v (u + y)^R
y^{2h - 2} dy - \frac{u^R}{(1 - 2h) \epsilon^{1 - 2h}} \right\} 
\]
if $h < 1/2$.} 
\begin{eqnarray*}
\frac{1}{v^{2h-1}} \int_0^v (u + y)^R y^{2h-2} dy & = & \sum_{r=0}^{\infty} 
\frac{\Gamma(R+1)}{r! \Gamma(R-r+1)} \frac{u^{R-r} v^r}{r+2h-1} \\
& = & \sum_{r=0}^{\infty} \frac{(-1)^r (u+v)^{R-r} \Gamma(R+1) \Gamma(2h-1)
v^r} {\Gamma(R-r+1) \Gamma(2h + r)}
\end{eqnarray*}
to rewrite eq.(\ref{6.17}) as 
\[
\psi(x) \simeq c_0 \sum_{r=0}^{\infty} \frac{\Gamma(R+1) \Gamma(2h)} {
\Gamma(R-r+1) \Gamma(2h+r)} \frac{(-2 \sqrt{\Lambda} a)^r}{(1-x)^{r+1}} 
\]
for $x$ close to 1. It then follows that 
\begin{eqnarray}
b_{n+f} \simeq c_0 \sum_{r=0}^{\infty} \frac{(-2 \sqrt{\Lambda} a)^r
\Gamma(2h)} {\Gamma (r + 2h)} \frac{\Gamma(R+1)}{r! \Gamma(R-r+1)} n^r
\label{6.19}
\end{eqnarray}
asymptotically for a large value of $n$. If $R$ were not a non-negative
integer, $b_{n+f}$ would be an infinite series in $r$ according to eq.(\ref%
{6.19}). This would then alter the asymptotic behavior of $a_{n+f}$ for large 
$n$. Hence $R$ has to be a non-negative integer, and eq.(\ref{6.19}) may be 
rewritten as 
\begin{eqnarray}
b_{n+f} \simeq c_0 \sum_{r=0}^R \frac{(-2 \sqrt{\Lambda} a)^r \Gamma(2h)} 
{\Gamma (r + 2h)} \left( 
\begin{array}{c}
R \\ 
r
\end{array}
\right) n^r,  \label{6.19a}
\end{eqnarray}
in perfect agreement with eq.(31) of Ref.\cite{0108149}. Moreover, the
transition amplitude at the continuum limit is finite and non-zero only if 
$E^{\prime}= 0$. Hence the energy spectrum at the continuum limit is also in
perfect agreement with the last displayed formula of section~5.1 of Ref.\cite
{0108149}. Consequently, the fractional sector model is exactly the same as
that of the $sl(2)$ gravity model at the continuum limit.

The transition amplitude at the continuum limit is then 
\begin{equation}
\tilde{G}(L,L^{\prime };T)=\frac{\sqrt{\Lambda }}{\sinh (\sqrt{\Lambda }T)}
e^{-\sqrt{\Lambda }(L+L^{\prime })\coth (\sqrt{\Lambda }T)}I_{2h-1}\left( 
\frac{2\sqrt{\Lambda LL^{\prime }}}{\sinh (\sqrt{\Lambda }T)}\right) .
\label{6.21}
\end{equation}
Eq.(\ref{6.21}) holds true even if $h = 1/2$ because the solution to 
eqs.(\ref{6.6.1}) and (\ref{6.7.1}) is continuous in $h$.  For the closed 
universe model with one particle, $h=1/2$ in eq.(\ref{6.21}) and the transition 
amplitude is independent of the value of $\xi $ as claimed in section~\ref{s2}. 
Moreover, the constraint $E^{\prime }=0$ together with eq.(\ref{6.0.1.1}) 
implies eq.(\ref{2.5}). For the closed universe model with two distinguishable 
bosons, $h = 1$ in eq.(\ref{6.21}) and the constraint $E' = 0$ implies 
eq.(\ref{4.10}).  In particular, eq.(\ref{2pdclosed}) holds true even if $K_1 
+ K_2 \neq 0$ as claimed.  For the closed universe model with two 
indistinguishable bosons, $h=1$ in eq.(\ref{6.21}) and the transition amplitude 
is independent of the value of $K$. The requirement $E^{\prime }=0$ and 
eq.(\ref{6.0.5}) lead to a constraint between the mobility of the bosons and 
their masses: 
\[
V-2K=\frac{3}{4}. 
\]

\section{Summary}

\la{s7}

We have demonstrated in this paper how to incorporate particles into
string bit models of quantum gravity. By considering the particles to be
represented by marked links, we obtain a simple model of quantum gravity in
which the energy of the links influences the evolution of the spatial slice
and vice versa.

We find that such models with one or two particles generally behave in the
continuum limit as open or closed universes without particles. A single
particle in closed universe behaves as an open empty universe, and a single
particle (or two distinguishable particles) in an open universe behaves as
an empty closed universe, provided certain constraints hold between the
various coupling parameters in the models. In particular, the masses and
mobilities of the particles in the various models are linearly related to
each other. Our most general result is given in eq.(\ref{6.21}), which is
the continuum limit of the fractional sector model, as it contains several
previous cases.  Note that eq.(\ref{6.21}), in which $h$ is continuous, is
also the transition amplitudes of some other quantum gravity models.  For 
example, this is the transition amplitude of a two-dimensional continuum 
quantum gravity model in the proper time gauge in which $h$ is determined
by the conformal anomaly and the winding modes around the boundaries
\cite{nakayama}.  Alternatively, if one considers a two-dimensional 
Lorentzian gravity model in the presence of small baby universes, one will
find that the transition amplitude is also given by eq.(\ref{6.21}) in 
which $h$ is associated with the density of baby universes \cite{dgk01}.

We have incorporated the quantum particles into our model by considering
homogeneous distributions, ie. equal superposition of the particle in all
possible states. A more general situation would be to consider arbitrary
superpositions of such states. Work on this problem is in progress.

\vskip 1pc \noindent {\Large \textbf{Acknowledgment} \vskip 1pc }

{\normalsize \noindent We thank A. Ambj{\o}rn, B. Durhuus, and T. P.
Devereaux for discussion. }

{\normalsize \vskip 1pc \noindent }{\Large \textbf{Appendix: Canonical
Commutation Relations} \vskip 1pc }

{\normalsize \noindent Let $M_{\mu_2}^{\mu_1}$ be the $(\mu_1, \mu_2)$-th
entry of the matrix $M$. If $M$ has only one row, we will write its entries
as $M^{\mu}$; if it has only one column, we will write its entries as $%
M_{\mu}$. The annihilation operators $a$, $b$, $b_1$, $b_2$, and $b_{12}$
which appear in this article are square matrices of order $N$. The
corresponding creation operators are $a^{\dagger}$, $b^{\dagger}$, $%
b_1^{\dagger}$, $b_2^{\dagger}$, and $b_{12}^{\dagger}$, respectively. The
annihilation operators $q$ and $\bar{q}$ are $N$-dimensional row and column
vectors, respectively. The corresponding creation operators $q^{\dagger}$
and $\bar{q}^{\dagger}$ are column and row vectors, respectively. They
satisfy the canonical commutation relations 
\begin{eqnarray*}
\left\lb a_{\mu_1}^{\mu_2}, a_{\mu_3}^{\dagger\mu_4} \right\rb & = &
\delta_{\mu_1}^{\mu_4} \delta_{\mu_3}^{\mu_2}, \\
\left\lb b_{\mu_1}^{\mu_2}, b_{\mu_3}^{\dagger\mu_4} \right\rb & = &
\delta_{\mu_1}^{\mu_4} \delta_{\mu_3}^{\mu_2}, \\
\left\lb b_{1\mu_1}^{\mu_2}, b_{1\mu_3}^{\dagger\mu_4} \right\rb & = &
\delta_{\mu_1}^{\mu_4} \delta_{\mu_3}^{\mu_2}, \\
\left\lb b_{2\mu_1}^{\mu_2}, b_{2\mu_3}^{\dagger\mu_4} \right\rb & = &
\delta_{\mu_1}^{\mu_4} \delta_{\mu_3}^{\mu_2}, \\
\left\lb b_{12\mu_1}^{\mu_2}, b_{12\mu_3}^{\dagger\mu_4} \right\rb & = &
\delta_{\mu_1}^{\mu_4} \delta_{\mu_3}^{\mu_2}, \\
\left\lb q^{\mu_2}, q^{\dagger}_{\mu_2} \right\rb & = &
\delta_{\mu_1}^{\mu_2}, \; \mbox{and} \\
\left\lb \bar{q}_{\mu_2}, \bar{q}^{\dagger\mu_1} \right\rb & = &
\delta_{\mu_1}^{\mu_2}.
\end{eqnarray*}
All other commutators vanish. }

\end{document}